\documentclass{ws-ijmpa}

\usepackage[super]{cite}
\usepackage{xcolor}
\usepackage{graphicx}
\usepackage{subcaption}
\usepackage[verbose,hypertexnames=false]{hyperref}
\hypersetup{colorlinks=false,allbordercolors=blue,pdfborderstyle={/S/U/W 1}}

\begin{document}

\markboth{Bence Szentkuti, Máté Csanád}{Hydrodynamic solutions with Lévy distributions}

%
\catchline{}{}{}{}{}
%

\title{Lévy-shaped femtoscopic sources from an exact solution of relativistic hydrodynamics}

\author{Bence Szentkuti\textsuperscript{1}, Máté Csanád\textsuperscript{2}}

\address{Department of Atomic Physics, ELTE Eötvös Loránd University,\\ Pázmány Péter sétány 1/A, Budapest, H-1117, Hungary\\
\textsuperscript{1}szentkuti10@student.elte.hu, \textsuperscript{2}csanad@elte.hu}

\maketitle

\begin{history}
\received{Day Month Year}
\revised{Day Month Year}
\accepted{Day Month Year}
\published{Day Month Year}
\end{history}

\begin{abstract}
An important goal of high-energy physics is to better understand the spacetime structure of particle emitting sources created in heavy-ion collisions. For a fraction of a second, a special form of matter, the so-called quark-gluon plasma (QGP), is created, which can be examined with the toolkit of femtoscopy. With the measurement of correlation functions it was determined that Lévy distributions describe the shape of hadron (pion) emission sources at freeze-out the most accurately; however, analytic solutions of hydrodynamics typically yield Gaussian sources, and power-law tails are usually attributed to the hadronic stage rather than to the fluid dynamics itself. In this article we showcase an exact solution of the equations of relativistic hydrodynamics, which aims to resolve this puzzle by providing Lévy distributions directly from hydrodynamics. A detailed analysis reveals whether the predictions of this solution are in accordance with observations; in particular, the effective temperatures from angle-averaged transverse mass distributions and the femtoscopic radii from two-particle momentum correlation functions are analyzed.
\end{abstract}

\keywords{femtoscopy; heavy-ion physics; Lévy distributions.}

\ccode{PACS numbers: 25.75.Gz, 25.75.Ld, 24.10.Nz, 47.75.+f}

\section{Introduction}
This paper focuses on the intersection of hydrodynamics and femtoscopy, and attempts to investigate the ``puzzle'' presented by the success of hydrodynamics, whose exact solutions typically predict Gaussian spatial distributions and which is not known to produce power-law tails by itself, and the observed non-Gaussian, Lévy-stable spatial emission distributions in femtoscopy, as detailed below. We start by providing a brief introduction to femtoscopy, focusing on the distributions and functions that are necessary to understand this paper. Subsequently, the basics of hydrodynamics are discussed, finishing with a summary of experimental observations and the emergence of Lévy distributions.

\subsection{Femtoscopy}	
We shall begin with the source function $S(x,p)$, which describes the phase-space probability density of single-particle emission. From the source function, all observable
distributions can be calculated. The integral
\begin{equation}
    N_1(p)=\int S(x,p)\mathrm{d}^4 x
    \label{eq:one_mom_dis}
\end{equation}
provides us the most general form of a single-particle momentum distribution (often called momentum spectrum).

Femtoscopy uses correlation functions to determine the structure of particle emitting sources. The two-particle momentum correlation function is defined as
\begin{equation}
    C_2(p_1,p_2)=\frac{N_2(p_1,p_2)}{N_1(p_1)N_1(p_2)},
\end{equation}
where $N_2$ is the two-particle momentum distribution. Let us introduce $q=p_1-p_2$ and $K=(p_1+p_2)/2$ variables. If $K\approx p_1\approx p_2$ then the
\begin{equation}
    C_2(q,K)=1+\frac{|\Tilde{S}(q,K)|^2}{|\Tilde{S}(0,K)|^2}
    \label{eq:corr_one}
\end{equation}
approximate formula can be derived, where $\Tilde{S}$ denotes the $x\rightarrow q$ Fourier transform. Let us furthermore introduce the autocorrelation of the source function
\begin{equation} 
    D(x,K)=\int S\left(\rho+x/2,K\right)S\left(\rho-x/2,K\right)\mathrm{d}^4\rho, 
    \label{eq:D_rho} 
\end{equation} 
called the (spatial) pair distribution, where $\rho$ is the average position and $x$ is the relative coordinate of the pair. We define the Fourier transform as
\begin{equation}
    \Tilde{f}(q)=\int f(x)e^{iqx}\mathrm{d}^4x.
\end{equation}
With this convention $\Tilde{S}(0,K)=N_1(K)$ in Eq.~\eqref{eq:one_mom_dis}, and the inverse transform carries the factor $(2\pi)^{-4}$, as in the Lévy distribution later in Eq.~\eqref{eq:levy}; in any case, a constant prefactor would cancel in the ratios of Eqs.~\eqref{eq:corr_one} and \eqref{eq:corr_final}. Furthermore, the substitution $x_{1,2}=\rho\pm x/2$ gives 
\begin{align} 
    |\Tilde{S}(q,K)|^2&=\int S(x_1,K)S(x_2,K)e^{iq(x_1-x_2)}\mathrm{d}^4x_1\mathrm{d}^4x_2\nonumber\\
    &=\int D(x,K)e^{iqx}\mathrm{d}^4x=\Tilde{D}(q,K), 
\end{align} 
i.e., the Fourier transform of the pair distribution equals the squared modulus of the Fourier transform of the source. With this, Eq.~\eqref{eq:corr_one} becomes
\begin{equation}
    C_2(q,K)=1+\frac{\Tilde{D}(q,K)}{\Tilde{D}(0,K)}.
    \label{eq:corr_final}
\end{equation}
The key of femtoscopy lies in Eq.~\eqref{eq:corr_final}: this means that by measuring momentum correlations ($C_2$) we can infer spatial distributions, in other words, the structure of the source can be revealed.

\subsection{Relativistic hydrodynamics}\label{relhydro_intro}
To model the spacetime evolution of the QGP in collisions relativistic hydrodynamics is used. Assuming a perfect fluid the energy-momentum tensor has the following definition:
\begin{equation}
    T^{\mu\nu}=(\epsilon+p)u^\mu u^\nu-pg^{\mu\nu},
\end{equation}
where $\epsilon$ is the energy density, $p$ is pressure and $g^{\mu\nu}=\text{diag}(1,-1,-1,-1)$ is the metric tensor. If there is a conserved charge, then let $n$ denote its number density. The two basic equations are continuity and energy-momentum conservation:
\begin{align}
    \partial_\mu(nu^\mu)&=0,\\
    \partial_\nu T^{\mu\nu}&=0.
\end{align}
To find an analytic solution, an equation of state is also required. For our work
\begin{equation}
    \epsilon=\kappa p
\end{equation}
is used, with the additional assumption that
\begin{equation}
    p=nT,\label{eq:idealgas}
\end{equation}
where $T$ is the temperature.

\subsection{Exact hydrodynamic solutions}\label{hydrosec}
Solving the aforementioned equations of hydrodynamics is a nontrivial task; here we focus on a class of 1+3D exact solutions, which have been shown to agree with measured spectra, flow and radii\cite{Csorgo:2003ry,Csanad:2009wc,Csanad:2014dpa}. In a fairly general form we first choose a scaling variable
\begin{equation}
    \Bar{s}=F(s_x,s_y,s_z),
\end{equation}
where $F$ is an arbitrary function of $s_{x,y,z}=r_{x,y,z}/t$ (or equivalently, $r_{x,y,z}/\tau$, with the appropriate transformation of $F$). Its comoving derivative then vanishes:
\begin{equation}
    u^\mu\partial_\mu\Bar{s}=0.
    \label{eq:comowing}
\end{equation}
It is also required to define a function $\nu(\Bar{s})$ that is positive, differentiable and normalized such that $\nu=1$ at the origin, $\mathbf{r}=0$. With these satisfied, a general class of solutions can be written as
\begin{align}
    u^\mu&=\frac{x^\mu}{\tau},\label{eq:Hubble-velocity}\\
    n(t,\mathbf{r})&=n_0\left(\frac{\tau_0}{\tau}\right)^3\nu(\Bar{s}),\\
    p(t,\mathbf{r})&=p_0\left(\frac{\tau_0}{\tau}\right)^{3+3/\kappa},\\
    T(t,\mathbf{r})&=T_0\left(\frac{\tau_0}{\tau}\right)^{3/\kappa}\frac{1}{\nu(\Bar{s})}\label{eq:temperature_sol},
\end{align}
where $\tau=\sqrt{t^2-\mathbf{r}^2}$ is the coordinate proper time, $n_0,p_0,T_0,\tau_0$ are constants and $p_0=n_0T_0$ holds; furthermore, $\kappa$ has to be constant as well. It is important to note that members of this family of solutions are accelerationless. For the scaling variable and the scaling function, the solutions of Refs.~\refcite{Csorgo:2003ry} and~\refcite{Csorgo:2001ru} choose the following ansatz:
\begin{align}
    \Bar{s}&=\frac{r_x^2}{\dot X_0^2 t^2}+\frac{r_y^2}{\dot Y_0^2 t^2}+\frac{r_z^2}{\dot Z_0^2 t^2},\\
    \nu(\Bar{s})&=e^{-b\Bar{s}/2},
\end{align}
thus exhibiting ellipsoidal symmetry. Here $\dot X_0,\dot Y_0,\dot Z_0$ describe the source expansion rate, while $b$ is a constant. 

\subsection{Observables}\label{observs}

In order to compare to measurements, observables have to be calculated. As a first step, one needs to fix the hadronic freeze-out (FO) hypersurface. It is straightforward to assume the FO happening at a constant proper time $\tau_0$, but in general a $H(\tau)$ FO probability density can be assumed as well. The source function, via the Boltzmann--Jüttner distribution and the Cooper--Frye formula, can then be calculated as
\begin{align}
    S(x,p)\mathrm{d}^4x=\mathcal{N}B(x,p)H(\tau)\mathrm{d}\tau\mathrm{d}^3\textstyle\Sigma_\mu(x)p^\mu,
\end{align}
where $\mathcal{N}=g/(2\pi)^3$ with $g$ the degeneracy factor, $B(x,p)=\exp\left(\mu_\text{ch}/T-p^\mu u_\mu/T\right)$ is the Boltzmann distribution with chemical potential $\mu_\text{ch}(x)$, $H(\tau)$ is the freeze-out proper-time distribution, and $\mathrm{d}^3\Sigma_\mu(x)p^\mu$ is the Cooper--Frye factor\cite{PhysRevD.10.186}. The fugacity follows from the number density of a Boltzmann gas: 
\begin{align}
    n&=g\,e^{\mu_\text{ch}/T}\int\frac{\mathrm{d}^3p}{(2\pi)^3}e^{-E/T} 
     =\frac{g}{2\pi^2}\,e^{\mu_\text{ch}/T}\,m^2T K_2(m/T)\\
    \Rightarrow&\quad \mathcal{N}e^{\mu_\text{ch}/T}=\frac{n}{4\pi m^2T K_2(m/T)}. 
\end{align} 
For sudden freeze-out, $H(\tau)=\delta(\tau-\tau_0)$, this yields 
\begin{align} 
    S(x,p)\mathrm{d}^4x=\frac{n(x)}{4\pi m^2T(x)K_2\left(m/T(x)\right)} 
    \exp\left[-\frac{p^\mu u_\mu}{T(x)}\right]\frac{p^\mu u_\mu}{u^0} 
    \delta(\tau-\tau_0)\mathrm{d}\tau\mathrm{d}^3x, 
    \label{eq:source_func} 
\end{align} 
where $K_2$ is the modified Bessel function of the second kind of order two. The overall normalization (set by $n_0$) does not affect any of the quantities studied below. 

The observables from this solution were compared to RHIC data\cite{Csanad:2009wc} and were found to be compatible with the measured spectra, flow and radii. The solution also predicts the $R^{-2}\sim m_T$ scaling behavior in radii, where $m_T=\sqrt{m^2+K_T^2}$ is the transverse mass and $K_T$ is the pair average momentum in the transverse plane. The pair distributions calculated from this solution approximately follow Gaussians, which present a puzzle when acknowledging the observed Lévy-stable distributions in femtoscopy, as also discussed in the next subsection. Exact hydrodynamic solutions usually predict Gaussian-shaped sources. Numerical fluid-dynamical calculations do yield non-Gaussian correlation functions, owing to the flow profile and the freeze-out geometry; power-law tails of the pair distribution, however, are usually attributed to resonance decays and hadronic rescattering\cite{Frodermann:2006sp,PhysRevC.64.014902,Kincses:2024lnv} rather than to the fluid stage itself, whose large-distance behavior has rarely been investigated explicitly. The origin of the observed Lévy-shaped sources is therefore a problem yet to be fully understood.

\subsection{Lévy distributions in measurements}

During the past decades, the analyses at SPS\cite{doi:10.1142/S0217751X24440032}, RHIC\cite{universe3040085,universe9070300,universe10030102,PHENIX:2017ino} and LHC\cite{universe9070318,CMS:2023xyd,Csanad:2024hva} have shown that pair distributions from Eq. (\ref{eq:D_rho}) are best described by Lévy distributions. In this article we will use the following definition for a spherically symmetric Lévy-stable distribution:
\begin{equation}
    \mathcal{L}(\alpha,R,\mathbf{r})=\frac{1}{(2\pi)^3}\int\mathrm{d}^3\mathbf{q}\,e^{i\mathbf{q}\cdot\mathbf{r}}\,e^{-\frac{1}{2}(qR)^\alpha}
    \label{eq:levy}
\end{equation}
where $q=|\mathbf{q}|$, $\alpha$ is called the Lévy index and $R$ is the Lévy scale parameter. For our further work $0<\alpha\leq2$ will be allowed, as for $\alpha>2$ the distribution is not non-negative, hence not a probability density. It is also worth noting that for $\alpha=2$ the distribution is a Gaussian and for $\alpha<2$ a power-law tail (or sometimes called Lévy tail) appears with a $-3-\alpha$ exponent. It has been suggested that Lévy distributions appear at least partly due to the phenomenon of Lévy walk\cite{Kincses:2024lnv}, encompassing elastic and inelastic scatterings, as well as resonance decays. This suggests that even if the fluid-dynamical stage produces sources without power-law tails, Lévy walks distort them to power-law tailed Lévy distributions.

\subsection{Source size scaling in femtoscopy and the Lévy-hydro puzzle}

Source size scaling is the experimentally observed behavior of the Lévy scale parameter, namely that $R^{-2}\sim m_T$ holds. As we have previously mentioned, this scaling property is predicted by hydrodynamics, in particular by exact solutions leading to Gaussian-shaped sources. In the case of measured data this $m_T$ scaling of $R$ is clearly observable with Lévy-distributed sources present\cite{PHENIX:2017ino}, independently of particle type (for the case of pions and kaons see Refs.~\refcite{PHENIX:2017ino} and~\refcite{Kovacs:2023qax}). This means that even if Lévy distributions are created by a Lévy walk after the hydro evolution, the hydro-based scaling of the HBT radii is retained after it, despite the vastly different resonance contributions and scattering cross sections relevant for each particle species. The ``puzzle'' ahead of us is now clear: Lévy distributions appear and the hydrodynamic scaling property derived from Gaussian exact solutions holds as well; thus, we must reconcile measurement and theory. So far, no definitive explanation has been found that could account for the non-Gaussian sources and scaling simultaneously.

\section{Hydrodynamic solution with Lévy distributions}

In this section we showcase a solution of the above-mentioned equations of hydrodynamics, capable of creating pair distributions following Lévy distributions. Firstly, however, let us give motivation for our approach and method.

\subsection{Approach and motivation}
Explaining the appearance of Lévy tails in pair distributions has been studied before. One explanation is that secondary decays\cite{PhysRevC.64.014902} or hadronic rescattering after freeze-out lead to power-law tails, a process sometimes called a Lévy walk\cite{Kincses:2024lnv}. However, it is unlikely that a hydrodynamical scaling is preserved after hadronic scattering and decays as well, especially given the very different related characteristics of various hadrons. Hence Lévy walk still does not fully explain the experimentally observed hydrodynamic scaling of femtoscopic radii, which was predicted for hydrodynamic sources without power-law tails\cite{PHENIX:2017ino}.

In this article we investigate another possible explanation, by searching for an exact, analytic hydrodynamic solution that results in a Lévy-distributed source and generally realistic observables. Specifically, our goal is to find a solution to the equations of relativistic hydrodynamics, which, coupled with a freeze-out condition, leads to $D(x,K)$ distributions exhibiting power-law tails. Such a solution might also provide the scaling properties of femtoscopic radii that are seen in measurements, resolving the related puzzle.

\subsection{Limitations of our exact hydrodynamical approach}
Before presenting our solution, we discuss the limitations of the underlying assumptions from Sections \ref{relhydro_intro}--\ref{observs}. Firstly, the perfect fluid picture neglects viscous corrections, thus simplifying the form of the energy-momentum tensor. Secondly, further idealization comes from the sudden freeze-out condition, in the form of instantaneous particle emission on the hypersurface of $\tau=\tau_0$; in reality, however, this is a more gradual process. Thirdly, the ideal gas equation of state we use in Eq. (\ref{eq:idealgas}) leaves out interaction effects, and it is most appropriate in the dilute, weakly interacting limit.

Furthermore, we must note that the solution class we employ in this paper is accelerationless; more concretely, we are working with a Hubble-flow assumption. This choice makes the solution analytically tractable but limits its quantitative applicability to the full dynamics of heavy-ion collisions, especially in the early stages of expansion.

Another characteristic of the solution class is that the number density $n$ and temperature $T$ cannot both decrease when the distance from the origin is increased, as $n\sim\nu\left(\Bar{s}\right)$ but $T\sim1/\nu\left(\Bar{s}\right)$, meaning that for non-constant $\nu\left(\Bar{s}\right)$ a spatially diverging density or temperature is inevitable. It is important to note, however, that the source function itself is still decreasing outwards, meaning that the asymptotic region, where these unphysical values arise, contributes only negligibly to particle emission at freeze-out.

Finally, we would like to emphasize the benefits of the assumptions and idealizations discussed previously. A strength of this approach arises from the fact that it is possible to derive an analytic relation between the initial conditions, the equation of state (EoS) and the observables. With these models the possibility of fruitful numerical analysis also arises. In later sections both aspects are demonstrated, as the asymptotic power-law behavior of the source is derived analytically, while the appearance of Lévy-stable distributions in the physically relevant spatial regime is confirmed numerically.

\subsection{The solution with Lévy-distributed source}
Our solution is a member of the class introduced in Section \ref{hydrosec}. Its key object is the following scaling variable:
\begin{equation}
    \Bar{s}=\frac{r_x^2+\epsilon_0^2\tau^2}{v_x^2\tau^2}+\frac{r_y^2+\epsilon_0^2\tau^2}{v_y^2\tau^2}+\frac{r_z^2+\epsilon_0^2\tau^2}{v_z^2\tau^2},
\end{equation}
where $v_x,v_y,v_z$ denote scale parameters and $\epsilon_0$ is introduced as a dimensionless regularization parameter\footnote{This parameter is not to be confused with energy density shown earlier.}. For the scaling function we use:
\begin{equation}
    \nu(\Bar{s})=\left(s_0/\Bar{s}\right)^\beta,
\end{equation}
where $s_0\equiv\Bar{s}|_{\mathbf{r}=0}$ is defined as the value of $\Bar{s}$ at the spatial origin, so that by construction $\nu(\mathbf{r}=0)=1$. For now we leave $\beta$ as a parameter for further analysis and only fix its value later.

In the isotropic case $v_{x,y,z}\equiv v$ used in our numerical study, $s_0=3\epsilon_0^2/v^2$, hence on the freeze-out hypersurface
\begin{align}
    \nu&=\left(1+\frac{r^2}{R_c^2}\right)^{-\beta},\\
    n&=n_0\left(1+\frac{r^2}{R_c^2}\right)^{-\beta},\\
    T&=T_0\left(1+\frac{r^2}{R_c^2}\right)^{\beta},\\ 
    R_c&\equiv\sqrt{3}\,\epsilon_0\tau_0.\label{eq:core_radius} 
\end{align}
The overall scale $v$ thus drops out (only the ratios $v_x:v_y:v_z$ matter in general), and $\epsilon_0$ sets the core radius $R_c$ of the density profile ($R_c\approx10.4$~fm for $\epsilon_0=0.75$, $\tau_0=8$~fm). The parameter $\epsilon_0$ also regularizes the $\left(\Bar{s}\right)^{-\beta}$ singularity that would otherwise appear at $r=0$. Since $(\tau_0/\tau)^{3/\kappa}=1$ at $\tau=\tau_0$, the value of $\kappa$ does not enter the source function either.

With these definitions not only does Eq.~\eqref{eq:comowing} hold, but the source function from Eq. (\ref{eq:source_func}) also exhibits the asymptotic behavior in the ultrarelativistic regime ($T\gg m$):
\begin{equation}
    S(x,p)\Big|_{r\gg R_c}\sim r^{-8\beta},\;\text{where}\;r=\sqrt{r_x^2+r_y^2+r_z^2}
    \label{eq:S_atinfty}
\end{equation}
in the isotropic case and for $\beta>1/2$. This asymptotic form follows directly from Eqs.~(\ref{eq:Hubble-velocity})--(\ref{eq:temperature_sol}) and (\ref{eq:source_func}). On the freeze-out hypersurface $t=\sqrt{\tau_0^2+r^2}$ and $u^\mu=(t,\mathbf{r})/\tau_0$. For $r\gg R_c$ one has $n\simeq n_0(R_c/r)^{2\beta}$ and $T\simeq T_0(r/R_c)^{2\beta}$; the temperature grows with $r$, so the ultrarelativistic limit $m^2K_2(m/T)\to2T^2$ is reached automatically. Furthermore, $p^\mu u_\mu=(Et-\mathbf{p}\cdot\mathbf{r})/\tau_0\simeq r(E-p\cos\theta)/\tau_0$, where $\theta$ is the angle between $\mathbf{p}$ and $\mathbf{r}$. Hence $p^\mu u_\mu/T\propto r^{1-2\beta}\to0$ for $\beta>1/2$, so the Boltzmann factor tends to unity, while $p^\mu u_\mu/u^0=E-\mathbf{p}\cdot\mathbf{r}/t\to E-p\cos\theta$. Putting these together (after the $\tau$ integration),
\begin{equation} 
    S(x,p)\simeq\frac{n}{8\pi T^3}\left(E-p\cos\theta\right) 
    =\frac{n_0}{8\pi T_0^3}\left(\frac{R_c}{r}\right)^{8\beta}\left(E-p\cos\theta\right) 
    \propto r^{-8\beta}. 
\end{equation} 
The exponent does not depend on the momentum, and the condition $\beta>1/2$ is exactly what is needed for the Boltzmann factor to approach unity.

\begin{figure}
\centerline{\includegraphics[width=10.0cm]{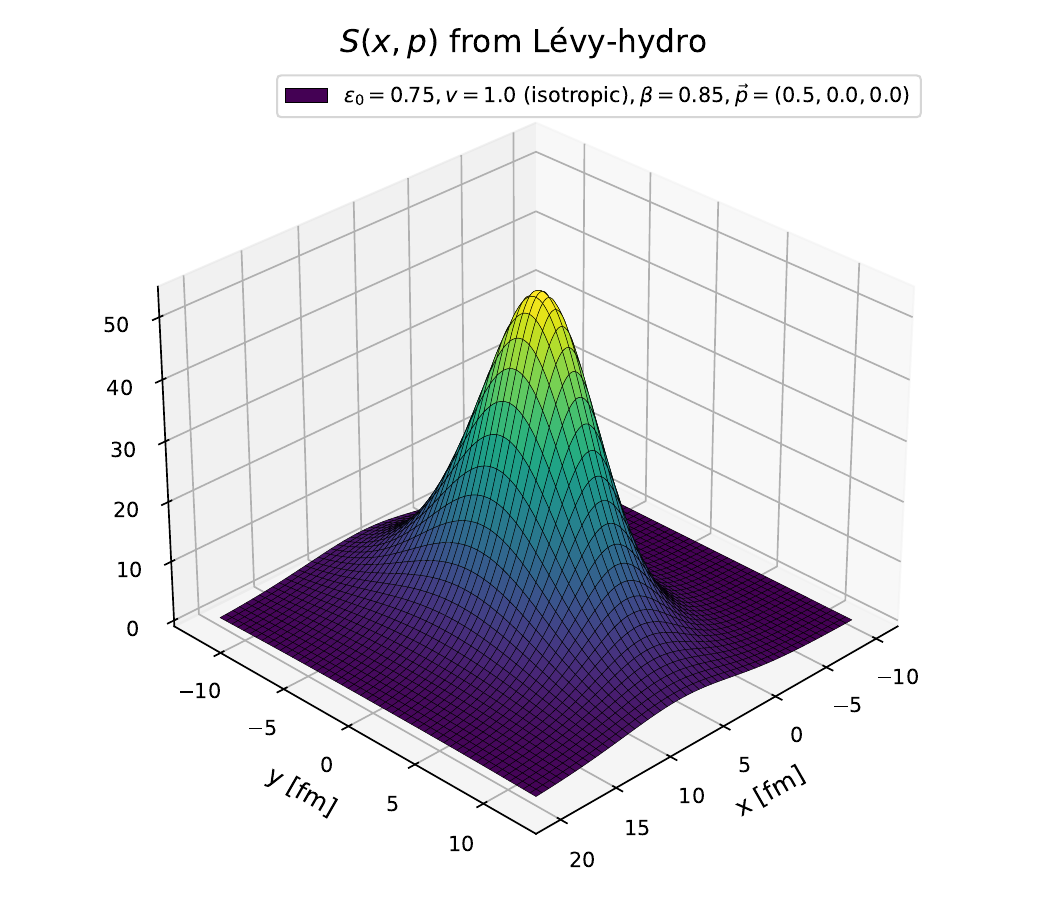}}
\caption{The source function calculated in the $x$--$y$ plane $(z = 0)$ at $\tau = \tau_0$, for $\mathbf{p} = (0.5, 0, 0)$ GeV, $\epsilon_0 = 0.75$, $v = 1$, $\beta = 0.85$ from the Lévy-hydro solution.}
\label{fig:sourcefunc2D}
\alttext{2D plot of the source function.}
\end{figure}

To check the validity of our solution we must determine suitable ranges of values for our parameters. It is worth noting that most of the parameters were found to have minor effect on our results, with the exceptions discussed below. The values chosen are presented in Table \ref{tab:params}, and stem from the literature\cite{Csanad:2009wc}, except for $\epsilon_0$ and $\beta$, for which reasonable ranges were investigated in detail.

\begin{table}[h!]
\tbl{Values of parameters used for the analysis ($c=1$).\label{tab:params}}
{\begin{tabular}{@{}ccc@{}} \toprule
Parameter & Notation & Value \\
\colrule
FO proper time & $\tau_0$ & 8.0 fm\\
Number density at FO & $n_0$ & 100.0 arb. units\\
Temperature at FO & $T_0$ & 0.17 GeV\\
Pion mass & $m$ & 0.14 GeV\\
Scale par. & $v$ & 1.0\\
Regularization par. & $\epsilon_0$ & 0.5--1.5\\
$\nu(\Bar{s})$ exponent & $\beta$ & 0.85\\
\botrule
\end{tabular}}
\end{table}

\begin{figure}[h]
\centerline{\includegraphics[width=7.0cm]{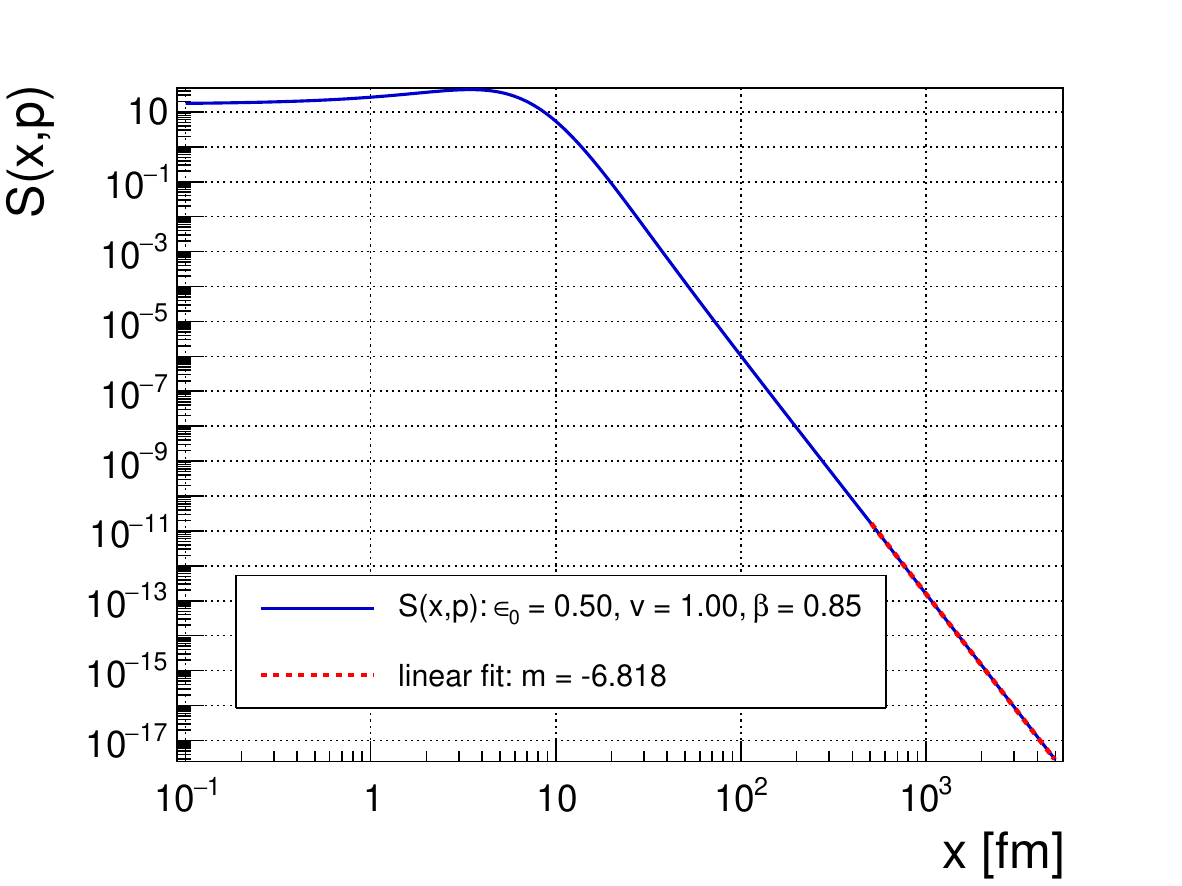}
\includegraphics[width=7.0cm]{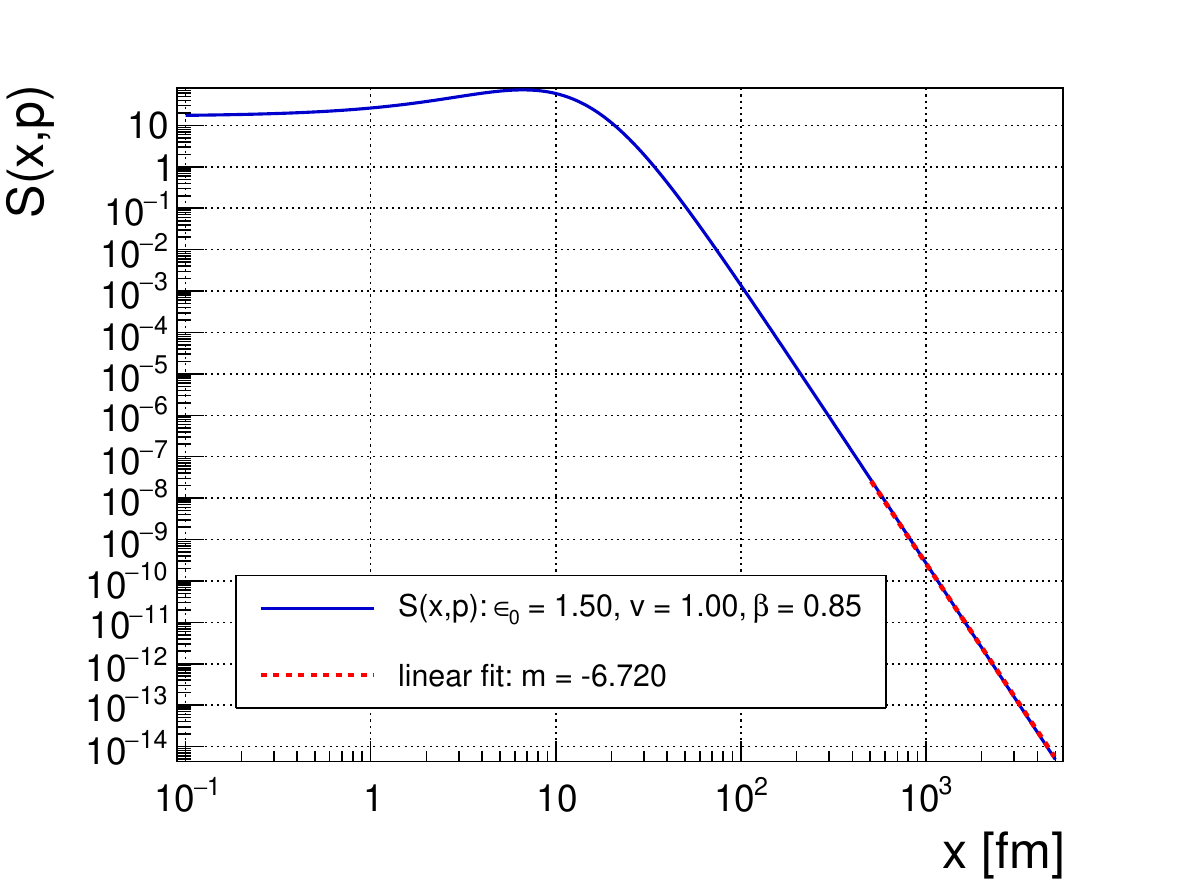}}
\caption{Two source functions generated from the Lévy-hydro solution along the positive $x$-axis, for $\epsilon_0=0.5$ (left) and $1.5$ (right), with $v=1$, $\beta=0.85$ and $\mathbf{p}=(0.5,0,0)$~GeV, as in Fig.~\ref{fig:sourcefunc2D}. To demonstrate the asymptotic behavior in Eq. (\ref{eq:S_atinfty}) a linear fit on log--log scale is carried out for larger $x$ values.}
\label{fig:sourcefuncs}
\alttext{Source functions calculated along the positive x-axis.}
\end{figure}

With these parameters set we can start our analysis by examining the properties of the source function. Fig.~\ref{fig:sourcefunc2D} shows the source function in the $x$--$y$ plane, while Fig.~\ref{fig:sourcefuncs} shows it along the $x$ axis on a log--log scale, illustrating that, as expected, power-law tails appear. Of all the parameters, $\epsilon_0$ has the greatest effect on the function's behavior, but only near the origin as it sets the core radius $R_c$ that we have introduced in Eq. (\ref{eq:core_radius}); for larger spatial coordinates, however, the predicted tails always prevail. The parameter $\beta$ is the main factor when it comes to setting the concrete value of the power-law decay, as seen in Eq. (\ref{eq:S_atinfty}). For our analysis, the value of $0.85$ was found to fulfill all our criteria, namely that it yielded a Lévy index $1<\alpha<2$, physical radii and effective temperatures near the measured values, all of which we discuss below. 

\begin{figure}[htbp]
    \centering
    \begin{subfigure}[b]{0.48\textwidth}
        \centering
        \includegraphics[width=\textwidth]{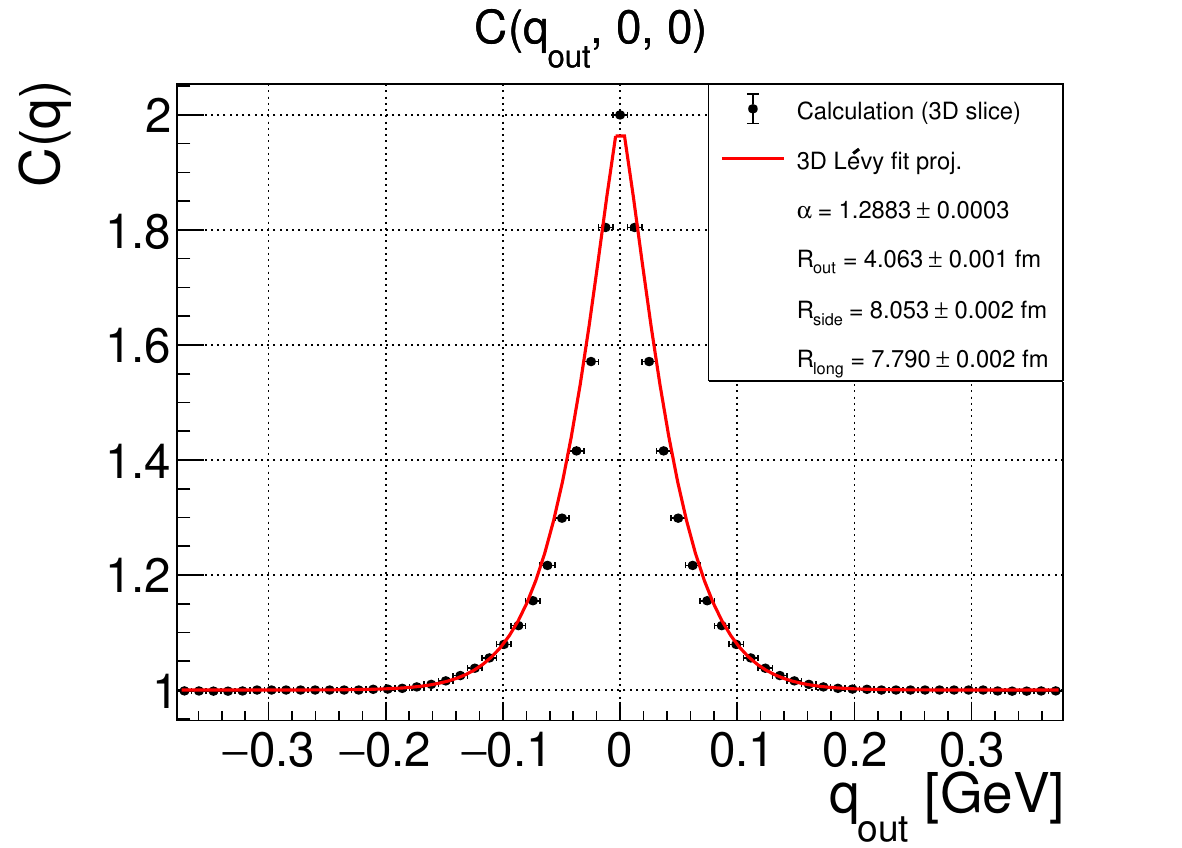}
    \end{subfigure}
    \hfill
    \begin{subfigure}[b]{0.48\textwidth}
        \centering
        \includegraphics[width=\textwidth]{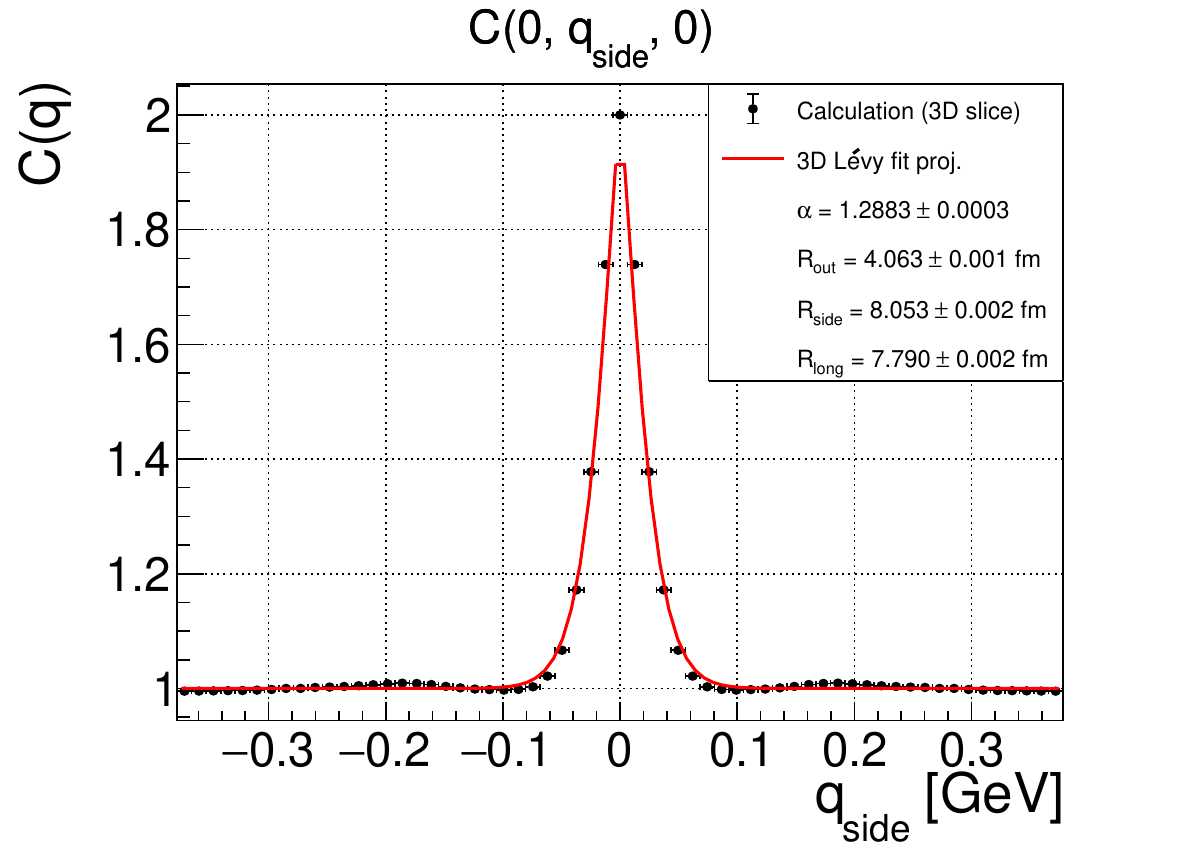}
    \end{subfigure}

    \vspace{0.4cm} 

    \begin{subfigure}[b]{0.48\textwidth}
        \centering
        \includegraphics[width=\textwidth]{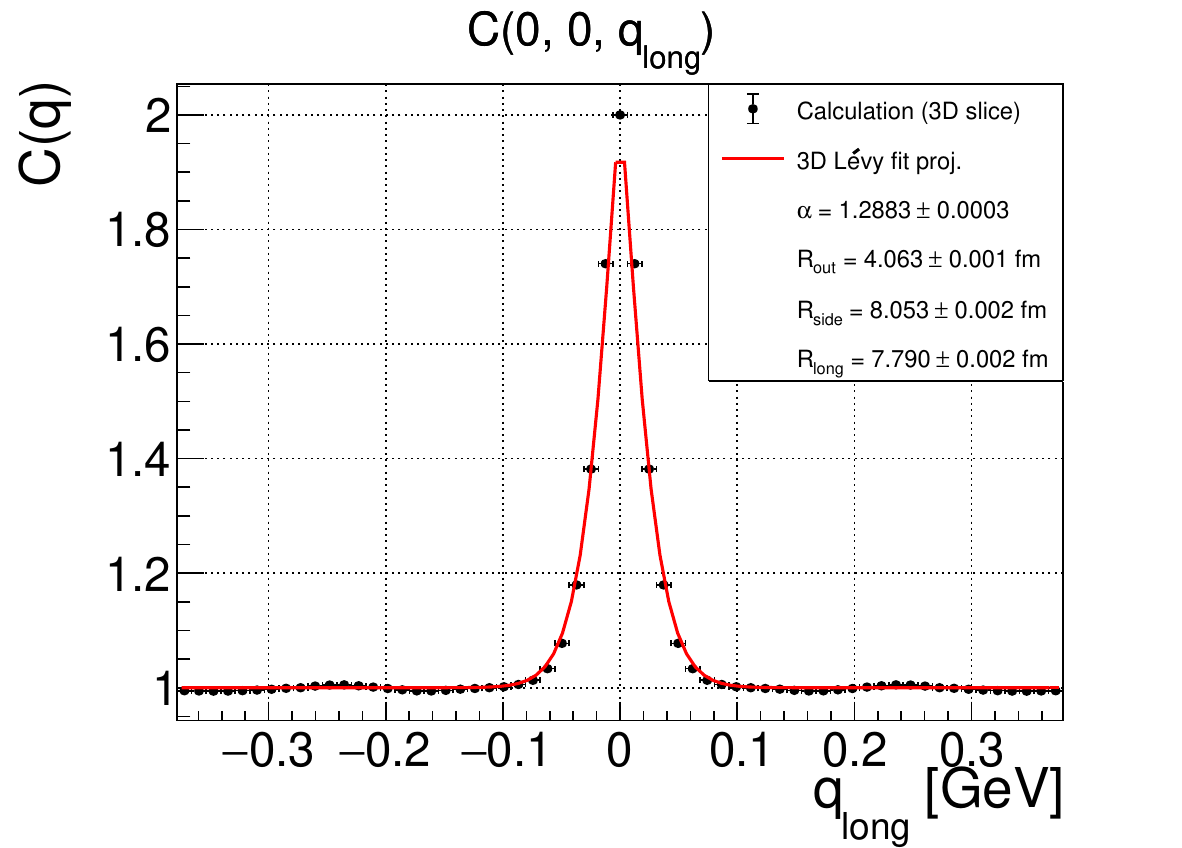}
    \end{subfigure}

    \caption{A calculated correlation function $C(\mathbf{q})$ plotted along the ``out'', ``side'' and ``long'' axes, calculated with $K_\text{out}=0.4$ GeV ($m_T\approx0.42$ GeV), $\epsilon_0=0.75$.}
    \label{fig:C_sidefits}
    \alttext{Calculated correlation functions.}
\end{figure}

\begin{figure}[htbp]
    \centering
    \begin{subfigure}[b]{0.48\textwidth}
        \centering
        \includegraphics[width=\textwidth]{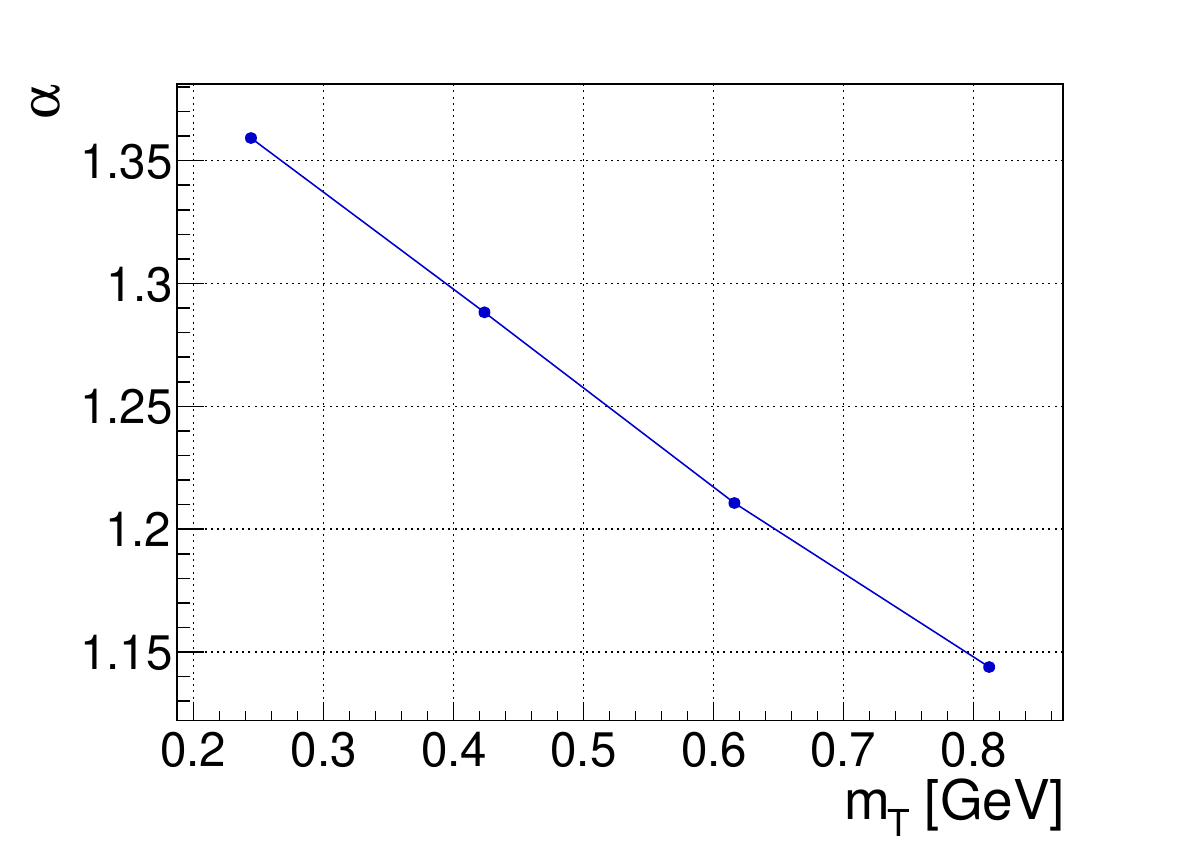}
    \end{subfigure}
    \hfill
    \begin{subfigure}[b]{0.48\textwidth}
        \centering
        \includegraphics[width=\textwidth]{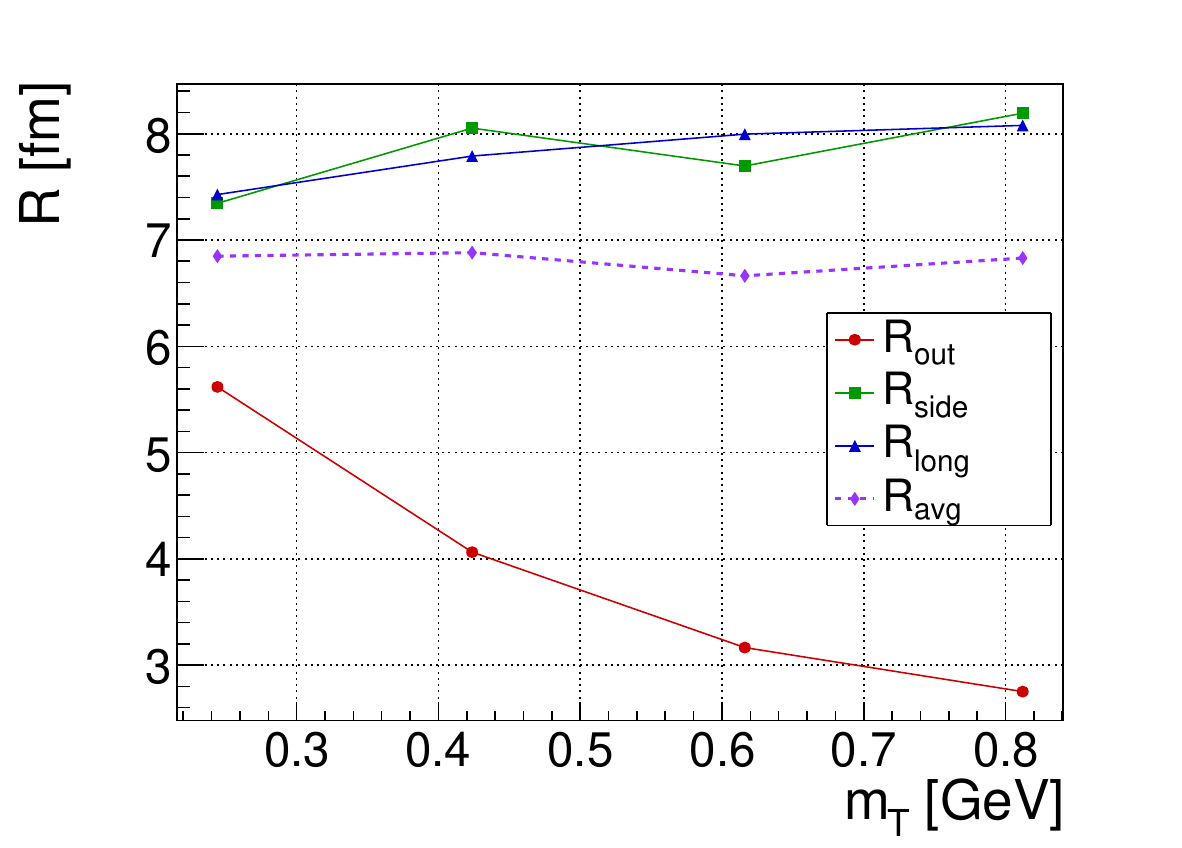}
    \end{subfigure}
    \caption{Parameters gathered from the fitting of correlation functions calculated with different $m_T$ values. The left plot shows $\alpha$, and the right one shows the radii with their $R_\text{avg}$ average value.}
    \label{fig:params_vs_mT}
    \alttext{Fitted parameters plotted against transverse mass.}
\end{figure}

We note here that the value of $\beta=0.85$ corresponds to a tail exponent of $-6.8$ in the source function (Fig. \ref{fig:sourcefuncs}), which is outside the range of Lévy-stable distributions with a tail exponent of $-3-\alpha$ (for $0<\alpha\leq2$). However, the correlation functions sample the intermediate spatial regime ($r\sim$ a few to a few tens of fm) and not the $r\rightarrow\infty$ limiting case. In fact, our later fits in Figs. \ref{fig:C_sidefits} and \ref{fig:params_vs_mT} confirm that in this regime the source is well approximated by a Lévy-stable distribution with an index $\alpha<2$. The term ``Lévy-distributed source'' should therefore be understood as referring to this effective description in the physically relevant regime, rather than the limiting case.

At large separations one particle of the pair is emitted from the core and the other one from the tail, so
\begin{equation}
    D(x,K)\simeq\left[S(x,K)+S(-x,K)\right]\int S(\rho,K)\,\mathrm{d}^4\rho,
\end{equation}
meaning the pair distribution inherits the $r^{-8\beta}$ tail of Eq.~\eqref{eq:S_atinfty}. Comparing it with the Lévy tail $r^{-3-\alpha}$, the asymptotic tail corresponds to $\alpha_\infty=8\beta-3$. A Lévy-compatible asymptotic tail ($0<\alpha_\infty\leq2$) is therefore obtained for $1/2<\beta\leq5/8$ (e.g., $\beta=0.55$ gives $\alpha_\infty=1.4$), while $\beta=0.85$ gives $\alpha_\infty=3.8$, i.e., a tail with finite variance. 

\begin{figure}[h]
\centerline{\includegraphics[width=7.0cm]{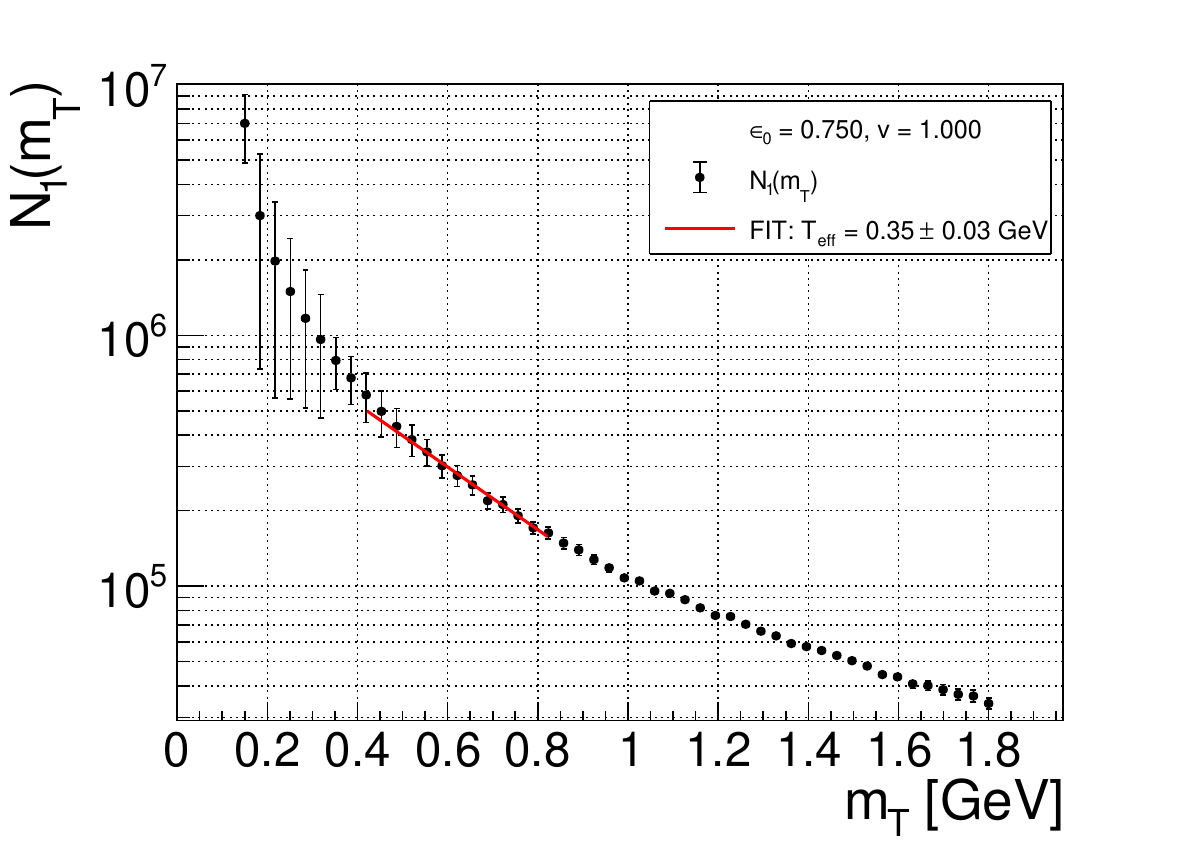}
\includegraphics[width=7.0cm]{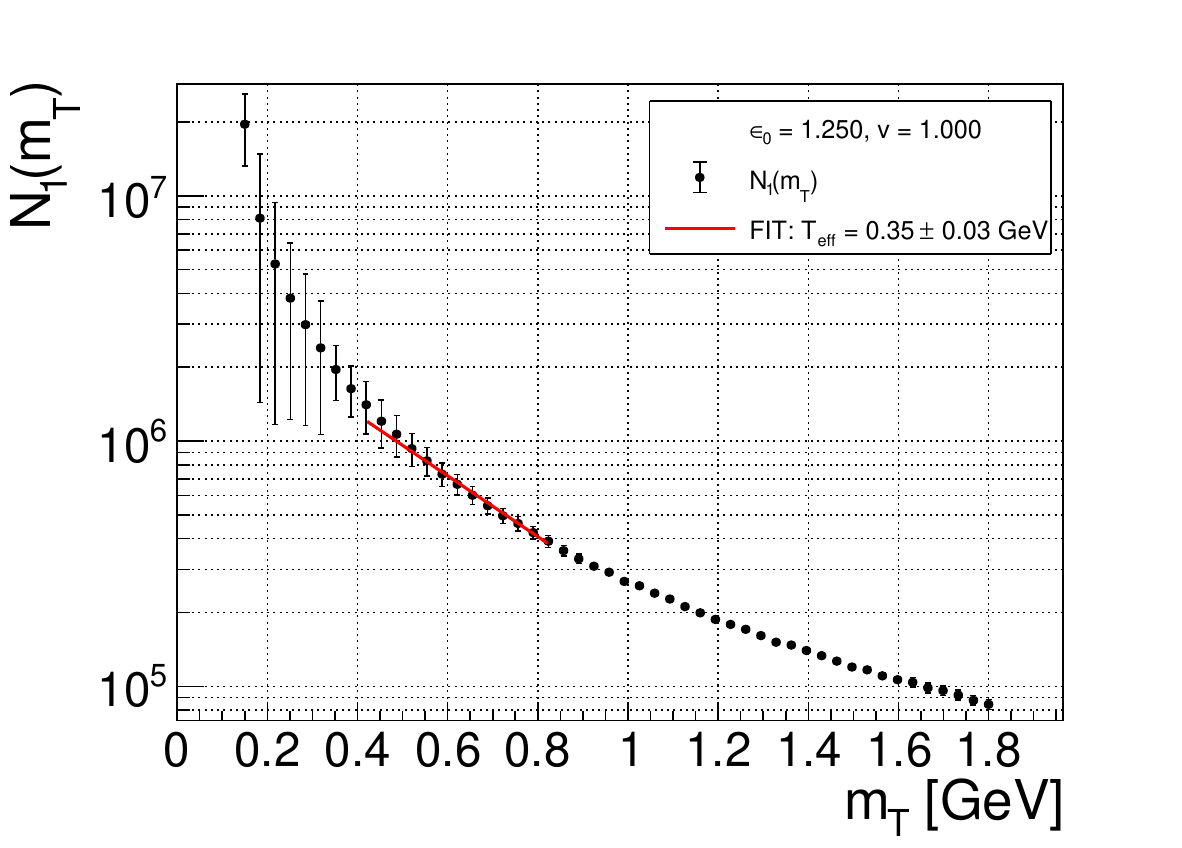}}
\caption{Calculated angle-averaged transverse mass distributions for $\epsilon_0=0.75$ (left) and $1.25$ (right), together with the fitted exponential of Eq.~(\ref{eq:fitted_expo_on_spectra}) and the resulting $T_\text{eff}$ values. Here $m_T$ on the horizontal axis denotes the single-particle transverse mass $m_t$.}
\label{fig:spectra}
\alttext{Figures showing the calculated spectra and effective temperature.}
\end{figure}

After the above analysis of the source function, we now examine momentum distributions calculated from our solution, specifically angle-averaged transverse mass distributions. These spectra were calculated using $\epsilon_0$ in the $[0.5,1.5]$ range, and were found to follow exponentials, agreeing with observations. We have chosen this range in $\epsilon_0$ as it sets the $R_c$ core radius in the range of $\sim$7--21 fm, which we deem physical. The calculation was carried out by first using Monte Carlo integration to determine the transverse momentum distribution according to Eq. (\ref{eq:one_mom_dis}), then the angle average was computed in the transverse plane using the trapezoidal rule. A simple exponential function in the form
\begin{equation}
    f(m_t,A,T_\text{eff})=A\cdot e^{-(m_t-m)/T_\text{eff}}
    \label{eq:fitted_expo_on_spectra}
\end{equation}
was fitted, where $m_t=\sqrt{p_t^2+m^2}$ is the single-particle transverse mass and $m=0.14$ GeV was used, assuming pions. The exponential was fitted to every spectrum in the $m_t\in[0.4,0.8]$ GeV range, where the main focus was on the extracted $T_\text{eff}$ effective temperature parameter, which in experiments is typically measured to be around $0.2$--$0.3$~GeV\cite{PHENIX:2003iij, Kurashvili:2015dit}. Two fitted spectra are shown in Fig.~\ref{fig:spectra} with different $\epsilon_0$ parameters. The errors of each point stem from the uncertainty in the integration, which we pass on to the Minuit2-based algorithm that fits the exponential from Eq. (\ref{eq:fitted_expo_on_spectra}). Our analysis showed that none of the fits produced effective temperatures in the desired range, but rather around $0.35$ GeV for all $\epsilon_0$, meaning we are above the desired $T_\text{eff}$ values.

For our final step, we chose, somewhat arbitrarily, the value $\epsilon_0=0.75$ (all values yield spectra with similar $T_\text{eff}$) and calculated the pair distribution $D(x,K)$ according to Eq.~\eqref{eq:D_rho}. To extract the $\alpha$ and $R$ parameters, we could try to directly match a Lévy-stable distribution to $D(x,K)$, however, taking the FFT and then fitting
\begin{equation}
    \mathcal{G}(\mathbf{q},\alpha,R_\text{out},R_\text{side},R_\text{long})=1+\exp\left[-\left(\frac{q_\text{out}^2R_\text{out}^2+q_\text{side}^2R_\text{side}^2+q_\text{long}^2R_\text{long}^2}{\hbar^2c^2}\right)^{\alpha/2}\right]
    \label{eq:corrrel_fitting}
\end{equation}
was computationally more efficient. We note here that in a more general case the correlation strength parameter $\lambda$ would also appear in $\mathcal{G}$; however, we are working with a pure hydrodynamical emission model, meaning $\lambda=1$ is fixed here.

The choice of reference frame here is not the laboratory frame. We utilize the Longitudinally Co-Moving System (LCMS) where $K_\text{long}=0$, so in the direction of the beam the momenta of the particle pair are of equal magnitude and are of opposite sign. It is also convenient to introduce the Bertsch--Pratt (BP) coordinates, with the ``out'' and ``side'' directions falling in the perpendicular (or transverse) plane of the ``long'' axis (beam direction). This frame is convenient for our analysis, as the average pair momentum vector takes the following form:
\begin{equation}
    \mathbf{K}=\left(K_\text{out},K_\text{side}=0, K_\text{long}=0\right).
\end{equation}
We note that the fitting function in Eq. (\ref{eq:corrrel_fitting}) already utilizes the BP coordinate system and the LCMS, furthermore the vector of $\mathbf{q}=\left(q_\text{out},q_\text{side},q_\text{long}\right)$ can be directly calculated from the lab frame components based on Ref.~\refcite{Kincses:2024lnv}. The FFT is taken over all four spacetime coordinates, including the time component. On the freeze-out hypersurface $\tau = \tau_0$, the time coordinate is $t = \sqrt{\tau_0^2 + r^2}$, which is correlated with $r$; this dependence is fully accounted for in the numerical evaluation of the pair distribution $D(x, K)$. The on-shell relation $q^0\approx\boldsymbol{\beta}\cdot\mathbf{q}=\beta_T q_\text{out}$, with $\boldsymbol{\beta}=\mathbf{K}/K^0$ and $\beta_T=K_\text{out}/K^0$, is used consistently when transforming between the lab frame and LCMS coordinates, following the prescription of Ref.~\refcite{Kincses:2024lnv}.


After clarifying our approach, we calculated the correlation functions and fitted them using Eq. (\ref{eq:corrrel_fitting}), resulting in figures such as Fig. \ref{fig:C_sidefits}. Our goal was not only to recover physical values for the $\alpha$ and radii, but to investigate their dependence on the $m_T$ transverse mass, specifically we were interested in whether the hydrodynamic scaling of radii is reconstructed by our solution. To test the latter, we calculated the correlation functions with different inputs in $K_\text{out}$ and fitted each of them, then plotted the parameters as a function of $m_T$ in Fig. \ref{fig:params_vs_mT}. For the concrete values of $K_\text{out}$ we have used 0.2, 0.4, 0.6, and 0.8~GeV, and for calculating $m_T$ the pion mass was utilized. The Lévy index $\alpha$ is in the correct range, but exhibits a clear $m_T$ dependence (from $1.35$ to $1.15$). The radii behave more interestingly: only $R_\text{out}$ decreases with $m_T$, while $R_\text{side}$ and $R_\text{long}$ increase slightly and are essentially equal within the numerical (Monte Carlo) fluctuations. The latter is a consequence of the geometry of the solution: our source is spherically symmetric, and in the LCMS and BP coordinates both the ``side'' and ``long'' directions are perpendicular to $\mathbf{K}$, so neither is preferred by the momentum dependence of the source. In the figure, we also plot the average radius, by which we mean:
\begin{equation}
    R_\text{avg}=\sqrt{\frac{1}{3}\left(R_\text{out}^2+R_\text{side}^2+R_\text{long}^2\right)}
\end{equation}
and this, in the case of our parameters, is approximately constant. Even if the gathered radii do not reproduce the well-known scaling property, we are still successful in the sense that their values are in the range given by measurements\cite{NA61SHINE:2023qzr,PhysRevC.110.064909}.

\section{Conclusions}

From the results presented above, we conclude that the presented Lévy-hydro solution is capable of creating the sought-after power-law tails in pair distributions. The introduced regularization parameter $\epsilon_0$ was found to influence the source function in the core region ($\sim$10 fm range); further out, however, the effect of $\beta$ dominates by setting the exact power of the decay. The analysis of the spectra and correlation functions revealed the successes and shortcomings of our work. The fit parameters are in or near the physical ranges, but do not fully agree with well-established results: the effective temperatures exceed the measured values by at least $0.05$~GeV, and the fitted radii do not reproduce the measured $m_T$ scaling.

Our results indicate that in the future a broader analysis of input parameters might yield values that align even better with measurements. If this path does not bring improvement, then it might be worthwhile to abandon the 1+3D accelerationless family of exact solutions, and consider other promising directions. These might include accelerating solutions or revisiting the hadronic Lévy walk picture.

In summary, our work may provide insight into solving the puzzle presented by the success of hydrodynamics and that of Lévy distributions.

\section*{Acknowledgments}
The authors are thankful to T. Csörgő, G. Kasza, and M. Nagy for the insightful discussions that led to these results. This project was supported by the NKFIH grant NKKP ADVANCED 152097.

\bibliographystyle{ws-ijmpa}
\bibliography{refs}

@article{Csanad:2009wc,
    author = "Csan\'ad, M\'at\'e and Vargyas, M\'arton",
    title = "{Observables from a solution of 1+3 dimensional relativistic hydrodynamics}",
    eprint = "0909.4842",
    archivePrefix = "arXiv",
    primaryClass = "nucl-th",
    doi = "10.1140/epja/i2010-10973-3",
    journal = "Eur. Phys. J. A",
    volume = "44",
    pages = "473--478",
    year = "2010"
}

@article{Csanad:2014dpa,
    author = "Csan\'ad, M\'at\'e and Szab\'o, Andr\'as",
    title = "{Multipole solution of hydrodynamics and higher order harmonics}",
    eprint = "1405.3877",
    archivePrefix = "arXiv",
    primaryClass = "nucl-th",
    doi = "10.1103/PhysRevC.90.054911",
    journal = "Phys. Rev. C",
    volume = "90",
    number = "5",
    pages = "054911",
    year = "2014"
}

@article{Csorgo:2003ry,
    author = {Cs{\"o}rg\H{o}, Tam{\'a}s and Csernai, L{\'a}szl{\'o} P{\'a}l and Hama, Yogiro and Kodama, Takeshi},
    title = "{Simple solutions of relativistic hydrodynamics for systems with ellipsoidal symmetry}",
    eprint = "nucl-th/0306004",
    archivePrefix = "arXiv",
    doi = "10.1556/APH.21.2004.1.8",
    journal = "Acta Phys. Hung. A",
    volume = "21",
    pages = "73--84",
    year = "2004"
}

@article{PhysRevD.10.186,
  title = {Single-particle distribution in the hydrodynamic and statistical thermodynamic models of multiparticle production},
  author = {Cooper, Fred and Frye, Graham},
  journal = {Phys. Rev. D},
  volume = {10},
  issue = {1},
  pages = {186--189},
  numpages = {0},
  year = {1974},
  month = {Jul},
  publisher = {American Physical Society},
  reportNumber = "Print-74-0742 (YESHIVA)",
  doi = {10.1103/PhysRevD.10.186},
  url = {https://link.aps.org/doi/10.1103/PhysRevD.10.186}
}

@article{PhysRevC.64.014902,
  title = "{Observing non-Gaussian sources in heavy-ion reactions}",
  author = {Brown, David A. and Danielewicz, Pawel},
  journal = {Phys. Rev. C},
  volume = {64},
  issue = {1},
  pages = {014902},
  numpages = {18},
  year = {2001},
  month = {Jun},
  publisher = {American Physical Society},
  doi = {10.1103/PhysRevC.64.014902},
  url = {https://link.aps.org/doi/10.1103/PhysRevC.64.014902}
}

@article{Kincses:2024lnv,
    author = "Kincses, D{\'a}niel and Nagy, M{\'a}rton and Csan{\'a}d, M{\'a}t{\'e}",
    title = "{L{\'e}vy walk of pions in heavy-ion collisions}",
    eprint = "2409.10373",
    archivePrefix = "arXiv",
    primaryClass = "nucl-th",
    doi = "10.1038/s42005-025-01973-x",
    journal = "Commun. Phys.",
    volume = "8",
    number = "1",
    pages = "55",
    year = "2025"
}

@article{NA61SHINE:2023qzr,
    author = "Adhikary, Haradhan and others",
    collaboration = "NA61/SHINE",
    title = "{Two-pion femtoscopic correlations in Be+Be collisions at $\sqrt{s_{\text {NN}}} = 16.84$ GeV measured by the NA61/SHINE~at CERN}",
    eprint = "2302.04593",
    archivePrefix = "arXiv",
    primaryClass = "nucl-ex",
    reportNumber = "FERMILAB-PUB-23-057-AD",
    doi = "10.1140/epjc/s10052-023-11997-8",
    journal = "Eur. Phys. J. C",
    volume = "83",
    number = "10",
    pages = "919",
    year = "2023"
}

@article{PhysRevC.110.064909,
  title = "{Centrality dependence of L\'evy-stable two-pion Bose-Einstein correlations in $\sqrt{{s}_{NN}}=200$ GeV $\text{Au}+\text{Au}$ collisions}",
  author = {Abdulameer N. J., Acharya and others},
  collaboration = {PHENIX},
  journal = {Phys. Rev. C},
  volume = {110},
  issue = {6},
  pages = {064909},
  numpages = {21},
  year = {2024},
  month = {Dec},
  publisher = {American Physical Society},
  doi = {10.1103/PhysRevC.110.064909},
  url = {https://link.aps.org/doi/10.1103/PhysRevC.110.064909}
}

@article{Csorgo:2001ru,
    author = "Cs{\"o}rg\H{o}, Tam{\'a}s",
    title = "{Simple solutions of fireball hydrodynamics for selfsimilar, ellipsoidal flows}",
    eprint = "hep-ph/0111139",
    archivePrefix = "arXiv",
    journal = "Acta Phys. Polon. B",
    volume = "37",
    pages = "483--494",
    year = "2006"
}

@article{PHENIX:2017ino,
    author = "Adare, Andrew and others",
    collaboration = "PHENIX",
    title = "{L{\'e}vy-stable two-pion Bose-Einstein correlations in $\sqrt{s_{NN}}=200$ GeV Au$+$Au collisions}",
    eprint = "1709.05649",
    archivePrefix = "arXiv",
    primaryClass = "nucl-ex",
    doi = "10.1103/PhysRevC.97.064911",
    journal = "Phys. Rev. C",
    volume = "97",
    number = "6",
    pages = "064911",
    year = "2018"
}

@article{PHENIX:2003iij,
    author = "Adler, Stephen S. and others",
    collaboration = "PHENIX",
    title = "{Identified charged particle spectra and yields in Au+Au collisions at $\sqrt{s_{NN}}=$200 GeV}",
    eprint = "nucl-ex/0307022",
    archivePrefix = "arXiv",
    doi = "10.1103/PhysRevC.69.034909",
    journal = "Phys. Rev. C",
    volume = "69",
    pages = "034909",
    year = "2004"
}

@Article{universe3040085,
AUTHOR = {Csanád, Máté},
TITLE = "{Lévy Femtoscopy with PHENIX at RHIC}",
JOURNAL = {Universe},
VOLUME = {3},
YEAR = {2017},
NUMBER = {4},
ARTICLE-NUMBER = {85},
URL = {https://www.mdpi.com/2218-1997/3/4/85},
ISSN = {2218-1997},
DOI = {10.3390/universe3040085}
}

@article{doi:10.1142/S0217751X24440032,
author = {P{\'o}rfy, Barnab{\'a}s},
title = "{Femtoscopy at NA61/SHINE using symmetric Lévy sources in central $^{40}$Ar+$^{45}$Sc from 40AGeV/c to 150AGeV/c}",
journal = {International Journal of Modern Physics A},
volume = {40},
number = {09},
pages = {2444003},
year = {2025},
doi = {10.1142/S0217751X24440032},
URL = {https://doi.org/10.1142/S0217751X24440032},
eprint = {https://doi.org/10.1142/S0217751X24440032}
}

@Article{universe9070300,
AUTHOR = {Mukherjee, Ayon},
TITLE = "{Kaon Femtoscopy with Lévy-Stable Sources from $\sqrt{s_{NN}}=$200 GeV Au+Au Collisions at RHIC}",
JOURNAL = {Universe},
VOLUME = {9},
YEAR = {2023},
NUMBER = {7},
ARTICLE-NUMBER = {300},
URL = {https://www.mdpi.com/2218-1997/9/7/300},
ISSN = {2218-1997},
DOI = {10.3390/universe9070300}
}

@Article{universe10030102,
AUTHOR = {Kincses, Dániel},
TITLE = "{Pion Interferometry with Lévy-Stable Sources in $\sqrt{s_{NN}}=$200 GeV Au + Au Collisions at STAR}",
JOURNAL = {Universe},
VOLUME = {10},
YEAR = {2024},
NUMBER = {3},
ARTICLE-NUMBER = {102},
URL = {https://www.mdpi.com/2218-1997/10/3/102},
ISSN = {2218-1997},
DOI = {10.3390/universe10030102}
}

@Article{universe9070318,
AUTHOR = {Kórodi, Balázs},
TITLE = "{Centrality-Dependent Lévy HBT Analysis in $\sqrt{s_{NN}}=$5.02 TeV PbPb Collisions at CMS}",
JOURNAL = {Universe},
VOLUME = {9},
YEAR = {2023},
NUMBER = {7},
ARTICLE-NUMBER = {318},
URL = {https://www.mdpi.com/2218-1997/9/7/318},
ISSN = {2218-1997},
DOI = {10.3390/universe9070318}
}

@article{Kovacs:2023qax,
    author = "Kov{\'a}cs, L{\'a}szl{\'o}",
    title = "{Charged Kaon Femtoscopy with L{\'e}vy Sources in $\sqrt{s_{\text{NN}}}$ = 200 GeV Au+Au Collisions at PHENIX}",
    eprint = "2307.09573",
    archivePrefix = "arXiv",
    primaryClass = "nucl-ex",
    doi = "10.3390/universe9070336",
    journal = "Universe",
    volume = "9",
    number = "7",
    pages = "336",
    year = "2023"
}

@inproceedings{Kurashvili:2015dit,
    author = "Kurashvili, Podist",
    collaboration = "ALICE",
    title = "{Photon and neutral pion production in pp and Pb-Pb collisions at LHC energies in the ALICE experiment}",
    booktitle = "{International School on High Energy Physics}: {Session C: Workshop in HEP}",
    eprint = "1511.00306",
    archivePrefix = "arXiv",
    primaryClass = "hep-ex",
    month = "11",
    year = "2015"
}

@article{CMS:2023xyd,
    author = "Tumasyan, Armen and others",
    collaboration = "CMS",
    title = "{Two-particle Bose-Einstein correlations and their L{\'e}vy parameters in PbPb collisions at $\sqrt{s_{\text{NN}}}$ = 5.02 TeV}",
    eprint = "2306.11574",
    archivePrefix = "arXiv",
    primaryClass = "nucl-ex",
    reportNumber = "CMS-HIN-21-011, CERN-EP-2023-094",
    doi = "10.1103/PhysRevC.109.024914",
    journal = "Phys. Rev. C",
    volume = "109",
    number = "2",
    pages = "024914",
    year = "2024"
}

@article{Csanad:2024hva,
    author = "Csan{\'a}d, M{\'a}t{\'e} and Kincses, D{\'a}niel",
    title = "{Femtoscopy with L{\'e}vy Sources from SPS through RHIC to LHC}",
    eprint = "2401.01249",
    archivePrefix = "arXiv",
    primaryClass = "hep-ph",
    doi = "10.3390/universe10020054",
    journal = "Universe",
    volume = "10",
    number = "2",
    pages = "54",
    year = "2024"
}

@article{Frodermann:2006sp,
    author = "Frodermann, Evan and Heinz, Ulrich and Lisa, Michael Annan",
    title = "{Fitted HBT radii versus space-time variances in flow-dominated models}",
    eprint = "nucl-th/0602023",
    archivePrefix = "arXiv",
    doi = "10.1103/PhysRevC.73.044908",
    journal = "Phys. Rev. C",
    volume = "73",
    pages = "044908",
    year = "2006"
}
\end{document}